\documentclass[%
 aip,
rsi,
 amsmath,amssymb,
reprint,%
]{revtex4-1}

\usepackage{xcolor}

\usepackage{graphicx}
\usepackage{dcolumn}
\usepackage{bm}

\usepackage[utf8]{inputenc}
\usepackage[T1]{fontenc}
\usepackage{mathptmx}

\graphicspath{{Figs/}}
\begin{document}

\preprint{AIP/123-QED}

\title[Hybrid magnetometer]{Spurious Ferromagnetic Remanence Detected by Hybrid Magnetometer}

\author{Giuseppe Bevilacqua}
\author{Valerio Biancalana}%
\email{valerio.biancalana@unisi.it}
\affiliation{Dept.\ of Information Engineering and Mathematics -  DIISM University of Siena - Via Roma 56, 53100 Siena,  Italy
}%
\author{Yordanka Dancheva}%
\author{Leonardo Stiaccini}
\author{Antonio Vigilante}
\affiliation{%
Dept.\ of Physical Sciences, Earth and Environment - DSFTA University of Siena - Via Roma 56, 53100 Siena,  Italy
}%

\date{\today}

\begin{abstract}
Nuclear magnetic resonance detection in ultra low field regime enables the measurement of different components of a spurious remanence in the polymeric material constituting the sample container.  A differential atomic magnetometer detects simultaneously the static field generated by the container and the time-dependent signal from the precessing nuclei. The nuclear precession responds with frequency shifts and decay rate variations to the container magnetization. Two components of the latter act independently on the atomic sensor and on the nuclear sample. A model of the measured signal allows a detailed interpretation, on the basis of the interaction geometry.
\end{abstract}

\maketitle

%


Optical atomic magnetometers (OAMs) find a variety of applications ranging from fundamental science \cite{afach_gnome_18, pendlebury_prd_15}  to security \cite{deans_apl_16}, nondestructive tests \cite{bevington_apl_18}, and nuclear magnetic resonance (NMR) detection in the regimes of zero and ultra-low-field \cite{tayler_rsi_17, biancalana_arnmrs_13} (ULF), including imaging \cite{savukov_jmr_14, oida_ieee_12}. 

OAM instrumentations are implemented in diverse designs, which lead to different complexity, sensitivity and robustness levels. Record sensitivities exceeding the fT/Hz$^{1/2}$ have been achieved with the spin-exchange-relaxation-free magnetometers, at expenses of a limited bandwidth and need of operating at vanishing field, so to require accurate shields and compensation systems. Less performing but more versatile implementations achieve 100~fT/Hz$^{1/2}$ in unshielded environment, and  miniaturized devices with analogous performance are nowadays developed \cite{schultze_oe_12, gerginov_josab_17}.
 
Excellent sensitivities are also obtained by radio-frequency  magnetometers, which are designed to detect high-frequency variations of the magnetic field. Recently a radio-frequency magnetometer was demonstrated to discriminate the polarization of the detected field \cite{gerginov_prappl_19}.  Other implementations reported in the literature, make use of cold atomic samples to improve the spatial resolution \cite{cohen_apl_19}.

In this paper we consider an OAM operating in unshielded environment with a bandwidth extending up to 200~Hz, which found recently application in ULF-NMR \cite{biancalana_jmr_09, biancalana_zulfJcoupling_jmr_16, biancalana_DH_jpcl_17} and imaging \cite{biancalana_IDEA_prappl_19} experiments.

We present a peculiar arrangement, where the ULF-NMR setup enables the characterization of weakly magnetized material on the basis of simultaneous analyses of the atomic and the nuclear precession signals. This arrangement leads to a hybrid measurement method, in which different components of the magnetization affect the nuclear and atomic spins, respectively.



\begin{figure}[ht]
   \centering
    \includegraphics [angle=0, width=  0.9 \columnwidth] {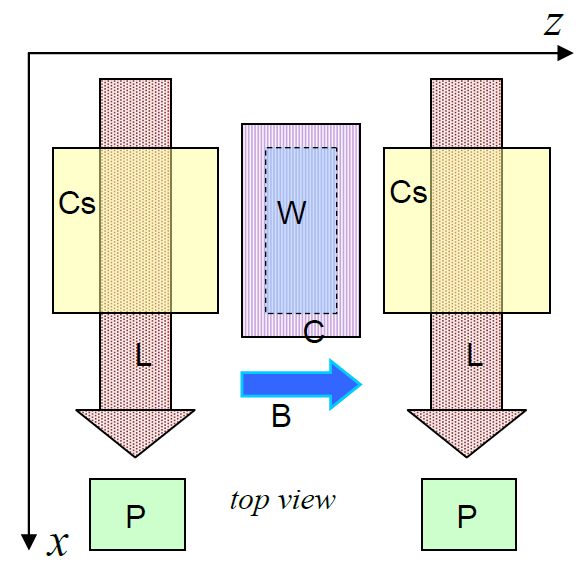}
  \caption{
	Top view of the dual magnetometric sensor. The sensors are made of Cs cells (Cs) interacting with  pump and probe laser beams (L) that co-propagate along the $x$ direction. A transverse static field at $\mu$T level is oriented along $z$ and polarimeters (P) measure the Faraday rotation of the probe beam polarization. For ULF-NMR measurements, a cartridge (C) containing water (W) is shuttled to the proximity of the sensor after having been magnetized. 
	}
		\label{fig:setupTop}
\end{figure}

The experimental setup (see Fig.\ref{fig:setupTop}) is built  around a dual OAM operating in a Bell \& Bloom configuration\cite{biancalana_apb_16} briefly described below.

The OAM uses Cs vapor that is optically pumped into a
stretched  state by means of laser
radiation at the milli-Watt level. This pump radiation is
circularly polarized and tuned to the Cs $D_1$ line. It is periodically tuned in resonance with the $F_g=3 \rightarrow F_e=4$  transition of the $D_1$ line set, to produce both light-narrowing \cite{sch_pra_11, biancalana_pra_16} due to strong hyperfine pumping to the $F_g=4$ ground state and  Zeeman pumping due to the weak (far detuned) resonances starting from the $F_g=4$ level excited with circular polarization.
A co-propagating weak (micro-Watt
level) and linearly polarized beam probes the time
evolution of the atomic state being tuned to the proximity
of the $F_g=4$ manyfold in the $D_2$ line. 

The periodicity of the pumping matches the precession frequency  around a magnetic field oriented transversely to the optical axis. Let $\omega_M$ and  $\Omega_L$ be the angular frequency of the laser modulation signal and of the atomic precession, respectively.
A scan of $\omega_M$ around $\Omega_L$ makes it possible to characterize the resonance profile, and linewidths as narrow as 25 Hz --set by spin exchange relaxation-- are recorded in operative conditions. 
The presence of buffer gas (23~Torr N$_2$) avoids line broadening due to Cs-wall collision and prevents radiation trapping phenomena.

Following interaction with the vapour, the pump radiation is stopped by an interference filter, and the polarization of the probe beams is analyzed by balanced polarimeters. During the measurements, $\omega_M$ is made resonant and kept constant. The magnetic field and its variation in time are extracted from the phase of the polarimetric signals.

The homogeneous B$_0$ field in which the sensors operate is obtained by
partially compensating the environmental field and is oriented along the $z$ axis.  $B_0$ has typical strengths of $1-10\, \mu$T. 

The system operates in an unshielded environment and the  external magnetic disturbances are first actively compensated \cite{biancalana_rsi_10, biancalana_prappl_19} and then cancelled by recording two signals differentially \cite{biancalana_apb_16}. To this end two identical Cs cells are used.

The sources of the measured signals can be modeled in terms of both static and precessing dipoles. It is important to consider the response of the magnetometer to field variations caused by such kind of sources, in dependence of their orientation.

\begin{figure}[ht]
   \centering
    \includegraphics [angle=0, width=  0.9 \columnwidth] {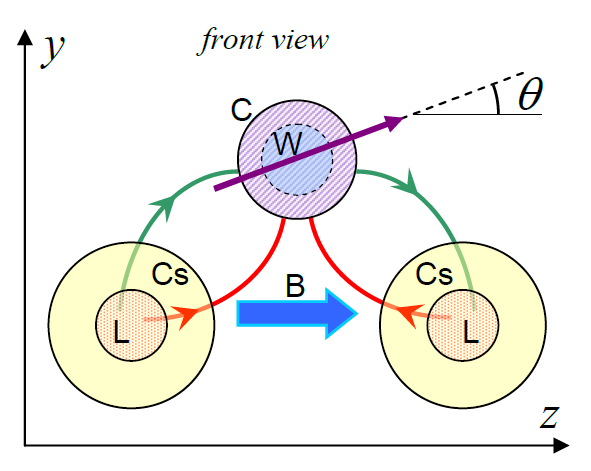}
  \caption{
		Front view of the measurement region. The differential response of the dual sensor is maximal for dipole sources oriented along $y$ (red arrows). The sample (C) undergo an unpredictable rotation by an angle $\theta$ during the transfer, so that the cartridge magnetization produces a static term proportional to $\sin \theta$. After an appropriate tipping pulse, the nuclear magnetization lies in the $xy$ plane and precesses around $z$, producing a time dependent signal. The precession frequency is set by the bias field superimposed to a perturbation due to the $z$ component of the cartridge magnetization.
	}
		\label{fig:setupFront}
\end{figure}

The scalar nature of the magnetometer makes it responding maximally to field-modulus variations, i.e. to  variations of the field component parallel to $\vec B_0$. Accordingly to the geometry --sketched in Fig.\ref{fig:setupFront}-- such variations are induced by dipolar sources located in the sample position and oriented along $y$ (red arrows in the figure), while dipolar sources oriented along $z$ produce (green arrows) smaller (second order response) field-modulus variations. Moreover, the latter gives a common mode signal that is compensated and cancelled. A similar condition would occur for an $x$ oriented dipole. In contrast, dipolar sources oriented along $y$ produce difference-mode field modulus variations ($\Delta B_{\mbox DM}$) to which the response of the differential setup is maximal.

The container of the NMR sample is a sealed polymeric cartridge. In this experiment, it is filled with  distilled water, other liquid substances can be analyzed as well, as previously reported \cite{biancalana_DH_jpcl_17, biancalana_zulfJcoupling_jmr_16}. 
The sample is remotely polarized (at about 1 T) in a Halbach magnetic array and pneumatically shuttled to the detection region \cite{biancalana_rsi_14}.

During the shuttling from the premagnetization Halbach array to the measurement region, the sample runs across a cylindrical pipe oriented along $x$. Unpredictable sample rotations by an angle $\theta$ may occur around the $x$ direction.

The nuclear magnetization follows adiabatically the external field so that it is oriented along $z$, when the sample reaches the measurement region.

The ULF-NMR signal is obtained by the application of an appropriate tipping pulse. In consequence of a $\pi/2$ pulse, the nuclear spins start precessing in the $xy$ plane and produce a time-dependent dipolar field that is optimally detected as a difference-mode signal.

Beside the signal oscillating at the nuclear precession frequency, a difference-mode static signal proportional to the $y$ component of the polymer magnetization is detected. Such static term appears with an amplitude dictated by the aleatory angle $\theta$, and is namely proportional to $\sin \theta$. In Fig.\ref{fig:setupFront}, the two cases of $\theta=0$ and $\theta=\pi/2$ are presented and they correspond to an additional static field represented by green and red arrows, respectively.


\begin{figure}[ht]
   \centering
    \includegraphics [angle=270, width=   \columnwidth] {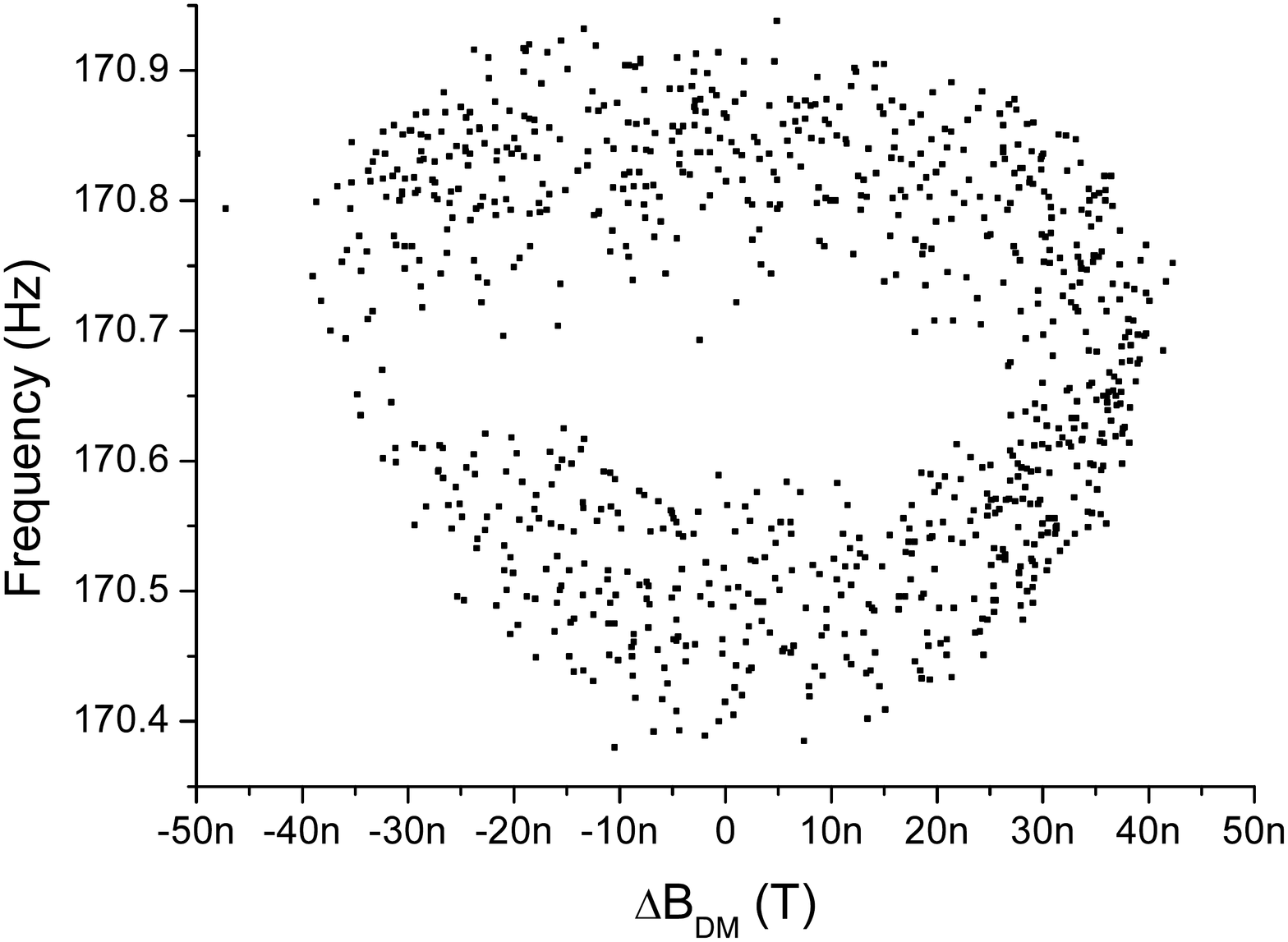}
  \caption{Nuclear precession frequency versus the difference-mode static field, estimated in a set of 880 measurements corresponding to a distribution of rotation angle $\theta$. The cartridge magnetization generates a difference-mode static field proportional to $\cos \theta$ (horizontal axis) and --correspondingly--  a shift of the nuclear precession frequency (vertical axis) that is proportional to $\sin \theta$. A remanence of $\sim 1 \mu$T is inferred from the measured static field. }
	\label{fig:wnvsb}
\end{figure}

\begin{figure}[ht]
   \centering
    \includegraphics [angle=270, width=  \columnwidth] {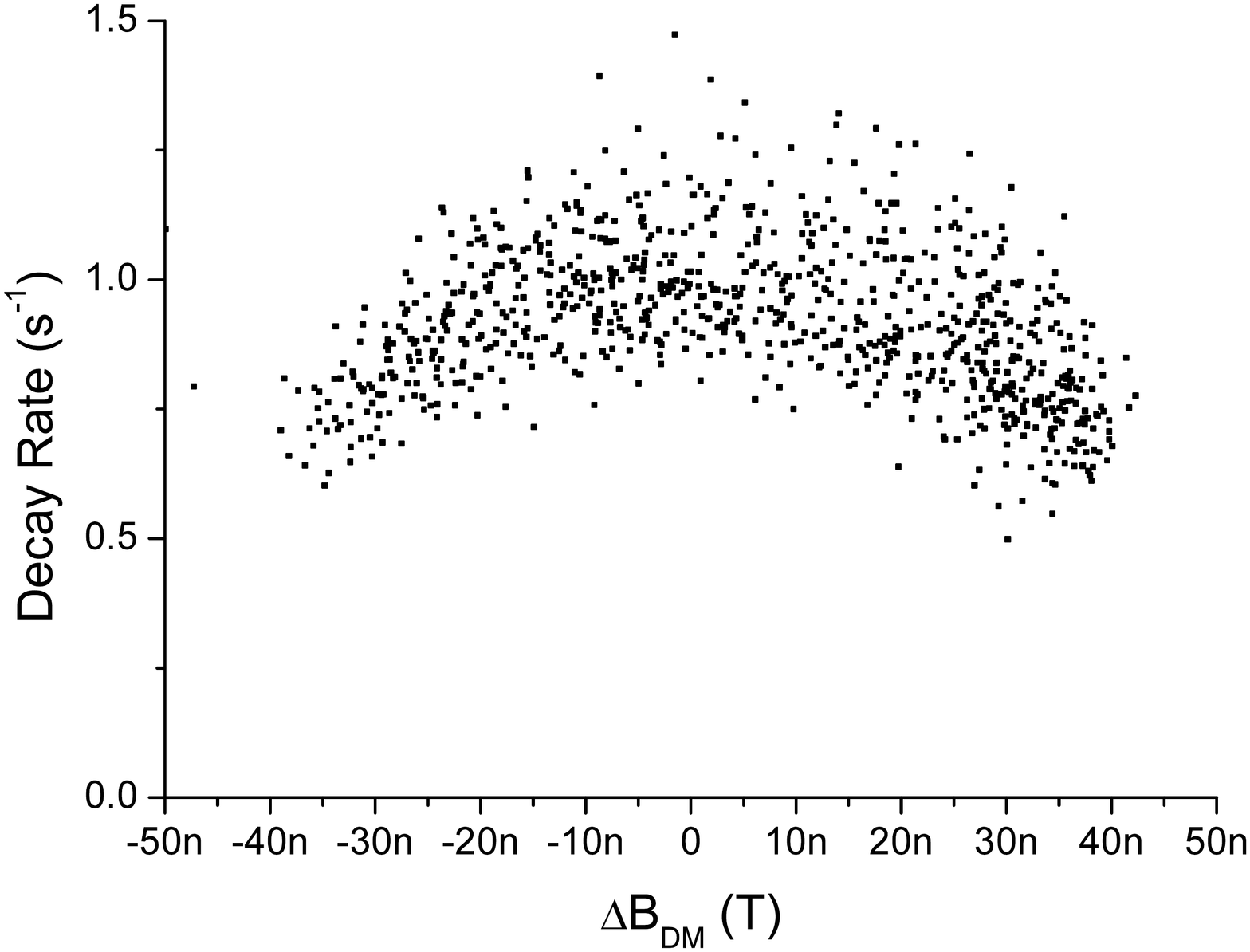}
  \caption{Estimated decay rate of the nuclear precession signal versus the difference-mode static field. The same measurements reported in Fig.\ref{fig:wnvsb} are analyzed. The spurious field modifying the nuclear precession frequency comes with some degree of inhomogeneity, so to accelerate the signal decay. The decay is faster in correspondence of $\theta$ values generating smaller smaller OAM static signal.}
		\label{fig:gnvsb}
\end{figure}

Beside the above mentioned static signal, the small but non-negligible magnetization $\vec M$ of the cartridge may cause a variation of the modulus of the field in which the protons precess. This variation  comes with some degree of inhomogeneity over the water volume, so that the nuclear spins experience slightly different local fields, with the emergence of an additional relaxation mechanism for the NMR signal.

Also in this respect, only the spurious field component along the static field $\vec B_0$ has first order effects, so that both the nuclear frequency shift and the decay rate are extreme when $\vec M$ is parallel (or antiparallel) to $\vec B$ while they are negligible when the two vectors are orthogonal (i.e. when $\theta=\pm \pi/2$ and maximal $\Delta B_{\mbox DM}$ is detected). 

A homogeneous magnetization $\vec M$ of an axially indefinite hollow cylinder, would not produce any field in the inner volume. In contrast, in our case, an internal field is present. This field is not homogeneous, has an average direction parallel to $\vec M$, and is caused by the finite length of the cylindrical distribution,  by the presence of end caps \cite{varga_ieee_98}, and possibly by inhomogeneities of the magnetization itself.

The Figs. \ref{fig:wnvsb} and \ref{fig:gnvsb} show the nuclear precession frequency and decay rate  over large set of ULF-NMR measurements. Both the quantities are estimated, trace by trace, making use of a Bertocco-Yoshida approach  \cite{bertocco_ieee_94,  duda_book_12}. It is worth noting that in our case the NMR signal is not expected to decay with a simple exponential law, because specific field inhomogeneities generated by the cartridge geometry drive the dephasing process. However, the development of a model that reproduces quantitatively the dynamics of the dephasing process is beyond the scope of this work. The mentioned  procedure, provides a qualitatively significant estimation of the decay rate, and clearly shows that shots where $\theta \approx 0$ (or $\pi$) result in larger $\gamma_2$ values, while smaller decay rates are obtained when the cartridge magnetization is perpendicular to the bias field ($\theta \approx \pm \pi/2$), so to maximize the static field term detected by the differential OAM.


In conclusion, we have discussed the use of an ULF-NMR setup to measure simultaneously the field variation caused by the cartridge magnetization both inside and outside the sample, based on the precession of protons and atoms, respectively. Such hybrid (atomic-nuclear) magnetometry provides -despite the scalar nature of the sensors- two independent and complementary kinds of measurements from which it is possible to infer both the modulus and the orientation of the magnetization.

The method can be of interest to characterize the spurious remanence of NMR samples. Sample ferromagnetic contamination constitutes a problem in ULF-NMR applications, whenever it increases relevantly the $\gamma_2$ parameter with respect to its intrinsic value. Moreover it reduces the accuracy in the determination of the nuclear precession frequency. And, finally, when cycled measurements  is necessary to improve the signal to noise ratio by trace averaging, the parasitic magnetization hinders the reproducibility of the nuclear precession frequency, so to cause  a $\gamma_2$ overestimation in the averaged signal.

\nocite{*}
\bibliography{aipsamp}

\end{document}